\documentclass[12pt]{iopart}
\usepackage{graphicx}
\usepackage[export]{adjustbox}
\usepackage[justification=centering]{caption}
\usepackage[font={footnotesize,it}]{caption}
\usepackage{cite}
\usepackage{iopams}  
\usepackage{soul}
\usepackage{bm}
\usepackage{hyperref}

\begin{document}

\title{Cavity-QED interactions of two correlated atoms}

\author{Saeideh Esfandiarpour}

\address{Department of Photonics, Faculty of Sciences and Modern Technologies, Graduate University of Advanced Technology, Kerman, Iran}

\author{Robert Bennett}

\address{Physikalisches Institut, Albert-Ludwigs-Universit\"at
Freiburg, Hermann-Herder-Str. 3, 79104 Freiburg, Germany}

\author{Hassan Safari}

\address{Department of Photonics, Faculty of Sciences and Modern Technologies, Graduate University of Advanced Technology, Kerman, Iran}

\author{Stefan Yoshi Buhmann}

\address{Physikalisches Institut, Albert-Ludwigs-Universit\"at
Freiburg, Hermann-Herder-Str. 3, 79104 Freiburg, Germany}

\address{Freiburg Institute for Advanced Studies,
Albert-Ludwigs-Universit\"at Freiburg, Albertstr. 19, 79104 Freiburg,
Germany}

\begin{abstract}

We consider the resonant van der Waals interaction between two correlated identical two-level atoms (at least one of which being excited) within the framework of macroscopic cavity quantum electrodynamics in linear, dispersing and absorbing media. The interaction of both atoms with the body-assisted electromagnetic field of the cavity is assumed to be strong. Our time-independent evaluation is based on an extended Jaynes-Cummings model. For a system prepared in a superposition of its dressed states, we derive the general form of the van der Waals forces, using a Lorentzian single mode approximation. We demonstrate the applicability of this approach by considering the case of a planar cavity and showing the position-dependence of Rabi oscillations. We also show that in the limiting case of weak coupling, our results reproduce the perturbative ones, for the case where the field is initially in vacuum state while the atomic state is in a superposition of two correlated states sharing one excitation.

\end{abstract}

%
\vspace{2pc}
\maketitle
%
%

\section{Introduction}
\label{1}
The van der Waals (vdW) force has been extensively studied for a long time. It has traditionally been derived from the interaction of two atoms with the continuous and unbounded spectrum of the electromagnetic field.
With this assumption, the atom-field interaction can be treated as perturbation. Being evaluated as the position dependent energy shift of two fluctuating dipoles interacting instantaneously, the vdW force in the weak coupling limit can be derived using time independent perturbation theory~\cite{buhreview2007}.\\
London first found this force between two ground-state atoms for small interatomic separations which are much smaller than the atomic transition wavelengths~\cite{London}. Casimir and Polder formulated it taking into account retardation due to the finite speed of light~\cite{CP}. Later the ideas have been extended to magnetic atoms~\cite{magneticCP} and also the effect of material environment was taken into account~\cite{materialmedia}. The theory had been generalized to identical atoms~\cite{identical1964,identical1966} and also to atoms in excited states, where conflicting oscillatory~\cite{excitedosci1965,excitedosci2003,excitedosci2015} and non-oscillatory~\cite{excitednonosci1993,excitednonosci1995} results, are predicted (for a review, see~\cite{buhreview2007,buhbookI}).\\

On the other side, the properties of atoms coupled to a discrete mode spectrum in a resonator-like structure are the 
subject of cavity quantum electrodynamics (cavity-QED). Historically, this field started with evaluating the effect of this coupling on the spontaneous emission probability of the atom, which was predicted to increase considerably due to the coupling with the cavity field~\cite{CQED1purcell}. \\
For strong interaction of one excited atom with an initially empty, un-damped discrete mode field, a model proposed by Jaynes and Cummings (JC model) predicts an oscillatory exchange of energy between the atomic and field systems that occurs at a rate $\Omega$, known as the Rabi frequency~\cite{jc}. The model involves calculating position dependent eigenenergies for the energy eigenstates of the coupled system (known as dressed states). Generalizing the static approach based on perturbation theory, the position dependent part of these eigenvalues can be related to dispersion forces~\cite{haroche1991trapping,englert1991reflecting}.
If $\Omega$ is much larger than the rate at which photons escape from the cavity, then the interaction is 
strong and the oscillation can be observed, otherwise the radiated photon may escape through the cavity walls before interacting with the atom~\cite{kaluzny1983observationCQED2}.\\

In this article, using a extension of JC model to absorbing cavities, we study the vdW force between two identical two-level atoms, which strongly interact with body-assisted electromagnetic field of the cavity.
We assume a dressed basis consisting of two correlated states in which the single excitation is shared between the atoms and the field. For two identical atoms the state that carries the excitation of two atoms can be assumed as a superposition of two states with only one atom being excited in each of them.
Following~\cite{strong2008}, the frequency dependence of the atomic responses to the field is taken into account in our calculations where the spectral structure of the field is assumed to be Lorentzian.

The article is organized as follows: first, we introduce the appropriate Hamiltonian for the system under consideration using a novel definition for the interaction Hamiltonian (\sref{2}); then in \sref{3} we evaluate the vdW force by combining the dressed state method, used for dealing with the strong atom-field coupling regime, with the primary idea of Casimir and Polder. We also evaluate the force in the weak coupling regime using leading order perturbation theory and show that it is in agreement with the appropriate limits of our results. An example of a planar cavity is provided in \sref{4} to demonstrate our model and finally a summary is given in \sref{5}.

\section{Atom-field interaction}
\label{2}

In order to derive the general form for the strong van der Waals (vdW) force, it is required to introduce the system Hamiltonian in the framework of macroscopic quantum electrodynamics.\\
Consider two neutral atoms (collection of charged particles) $\mathrm{A}$ and $\mathrm{B}$ positioned at $\mathbf{r}_{\mathrm{A}}$ and $\mathbf{r}_{\mathrm{B}}$, within an arbitrary arrangement of dispersing and absorbing magneto-electric bodies. The atoms are considered to interact strongly with the body-assisted electromagnetic field. The total Hamiltonian describing such a system in the multipolar coupling scheme reads

\begin{equation}
\label{E1}
\hat{H}=\hat{H}_{\mathrm{F}}+\sum_{\mathrm{A}'=\mathrm{A},\mathrm{B}}\left(\hat{H}_{\mathrm{A}'}+\hat{H}_{\mathrm{A}'\mathrm{F}} \right),
\end{equation}
where 
\begin{equation}
\label{E2}
\hat{H}_{\mathrm{A}'}=\sum_{n}E_{\mathrm{A}'}^n |n_{\mathrm{A}'}\rangle\langle n_{\mathrm{A}'}|
\end{equation}
is the internal Hamiltonian of atom $\mathrm{A}'$, with $E_{\mathrm{A}'}^n$ and $|n_{\mathrm{A}'}\rangle$ indicating the unperturbed eigenvalues and eigenstates, respectively. Considering the simple case of two identical two-level atoms with only two independent states $|0\rangle$ and $|1\rangle$, and discarding a state-independent constant, $\hat{H}_{\mathrm{A}'}$ reduces to
\begin{equation}
\label{E3}
\hat{H}_{\mathrm{A}'}=\frac{\hbar}{2}\omega_{\mathrm{A}'}^{10}\hat{\sigma}_{\mathrm{A}'}^{z},
\end{equation}
with  $\omega_{\mathrm{A}'}^{10}=(E_{\mathrm{A}'}^1-E_{\mathrm{A}'}^0)/\hbar$ and $\hat{\sigma}_{\mathrm{A}'}=|1_{\mathrm{A}'}\rangle\langle 1_{\mathrm{A}'}|-|0_{\mathrm{A}'}\rangle\langle 0_{\mathrm{A}'}|$.
\\

In order to introduce $\hat{H}_\mathrm{F}$, we use bosonic operators $\hat{\mathbf{f}}_\lambda^\dagger(\mathbf{r},\omega)$ and $\hat{\mathbf{f}}_\lambda(\mathbf{r},\omega)$ as the creation and annihilation operators of the body-assisted field excitation obeying the commutation relations
\begin{equation}
\label{E4}
[\hat{\mathbf{f}}_\lambda(\mathbf{r},\omega),\hat{\mathbf{f}}_\lambda'^\dagger(\mathbf{r}',\omega')]=\bm{\delta}(\mathbf{r}-\mathbf{r}')\delta(\omega-\omega')\delta_{\lambda \lambda'}
\end{equation}
and
\begin{eqnarray}
\label{E5}
[\hat{\mathbf{f}}_\lambda(\mathbf{r},\omega),\hat{\mathbf{f}}_\lambda'(\mathbf{r}',\omega')]=[\hat{\mathbf{f}}_\lambda^\dagger(\mathbf{r},\omega),\hat{\mathbf{f}}_\lambda'^\dagger(\mathbf{r}',\omega')]=\mathbf{0} .
\end{eqnarray}
 The field Hamiltonian then reads
\begin{eqnarray}
\label{E6}
\hat{H}_{\mathrm{F}}=\hbar \sum_{\lambda=e,m} \int d^3\mathbf{r} \int^{\infty}_0 d\omega \hspace{0.1cm} \omega \hspace{0.1cm}\hat{\mathbf{f}}^\dagger_\lambda(\mathbf{r},\omega)\cdot \hat{\mathbf{f}}_\lambda(\mathbf{r},\omega).
\end{eqnarray}

Now, having defined the atom and field Hamiltonians, we need to specify their interaction. In the electric dipole approximation, the multipolar Hamiltonian describing the interaction of an atom with the field can be written as
\begin{eqnarray}
\label{E7}
\hat{H}_{\mathrm{A}'\mathrm{F}}=-\hat{\mathbf{d}}_{\mathrm{A}'}\cdot  \hat{\mathbf{E}}(\mathbf{r}_{\mathrm{A}'}),
\end{eqnarray}
where for two-level atoms, the electric dipole moment is defined as
\begin{eqnarray}
\label{E8}
\hat{\mathbf{d}}_{\mathrm{A}'}=\mathbf{d}_{\mathrm{A}'}^{01} \hat{\sigma}_{\mathrm{A}'}+ \mathrm{H.c.},
\end{eqnarray}
with $\mathbf{d}_{\mathrm{A}'}^{mn}=\langle m_{\mathrm{A}'} |\hat{\mathbf{d}}_{\mathrm{A}'} |n_{\mathrm{A}'} \rangle$,  $\hat{\sigma}_{\mathrm{A}'}=|0_{\mathrm{A}'}\rangle\langle 1_{\mathrm{A}'}|$) and
$\hat{\mathbf{E}}$ being the electric field operator.

In order to facilitate giving the electric field as a linear combination of fundamental 
operators $\hat{\mathbf{f}}_\lambda$ and $\hat{\mathbf{f}}^\dagger_\lambda$, we employ the auxiliary tensors $\mathbf{G}_e$ and $\mathbf{G}_m$
defined in terms of the Green's tensor $\mathbf{G}$ as \cite{dung2003}
\begin{eqnarray}
\label{E9}
\mathbf{G}_e(\mathbf{r},\mathbf{r}',\omega)=i \frac{\omega^2}{c^2}\sqrt{\frac{\hbar}{\pi \varepsilon_0}\mathrm{Im} \varepsilon(\mathbf{r}',\omega)} \mathbf{G}(\mathbf{r}, \mathbf{r}', \omega),
\end{eqnarray}
\begin{eqnarray}
\label{E10}
\mathbf{G}_m(\mathbf{r},\mathbf{r}',\omega)= i \frac{\omega}{c}\sqrt{-\frac{\hbar}{\pi \varepsilon_0}\mathrm{Im} \mu^{-1}(\mathbf{r}',\omega) }
[\nabla'\times \mathbf{G}(\mathbf{r}',\mathbf{r},\omega)]^\top.
\end{eqnarray}
The Green's tensor obeys the differential equation 
\begin{eqnarray}
\label{E11}
\nabla\times \mu^{-1}(\mathbf{r},\omega) \nabla\times\mathbf{G}(\mathbf{r},\mathbf{r'},\omega) -\frac{\omega^2}{c^2}\varepsilon(\mathbf{r},\omega)\mathbf{G}(\mathbf{r},\mathbf{r'},\omega)=\bm{\delta}(\mathbf{r}-\mathbf{r'}),
\end{eqnarray}
subject to the boundary conditions imposed by the particular arrangement of magneto-dielectric bodies at hand. It also satisfies a useful integral relation
\begin{eqnarray}
\label{E12}
\sum_{\lambda=e,m} \int d s\, \mathbf{G}_\lambda(\mathbf{r},\mathbf{s},\omega).\mathbf{G}_\lambda^{\star T}(\mathbf{r}',\mathbf{s},\omega)
=\frac{\hbar\mu_0}{\pi}\omega^2 \mathrm{Im} \mathbf{G}(\mathbf{r},\mathbf{r}',\omega).
\end{eqnarray}
In Eqs.~(\ref{E9})-(\ref{E12}), $\varepsilon$ and $\mu$ are the position- and frequency-dependent relative electric permittivity and magnetic permeability
 of the surrounding media. 
With these definitions, the electric field can be written as follows
\begin{eqnarray}
\label{E13}
\hat{\mathbf{E}}(\mathbf{r})= \sum_{\lambda=e,m}\int_0^\infty d\omega \int d^3 r' \mathbf{G}_\lambda(\mathbf{r},\mathbf{r'},\omega)\cdot \hat{\mathbf{f}}_\lambda (\mathbf{r'},\omega)+\mathrm{H.c.}\,.
\end{eqnarray}

Using the field expansion (\ref{E13}) together with Eq. (\ref{E8}), the interaction Hamiltonian (\ref{E7}) reads
\begin{eqnarray}
\label{E14}
\fl \hat{H}_{\mathrm{A}'\mathrm{F}}=-\sum_{\lambda=e,m}\int_0^\infty\!\! d\omega \int \!\!d^3 r (\mathbf{d}_{\mathrm{A}'}^{01} \hat{\sigma}_{\mathrm{A}'}+\mathbf{d}_{\mathrm{A}'}^{10}
 \hat{\sigma}_{\mathrm{A}'}^\dagger) \mathbf{G}_\lambda(\mathbf{r}_{\mathrm{A}'},\mathbf{r},\omega)\cdot \hat{\mathbf{f}}_\lambda (\mathbf{r},\omega)+\mathrm{H.c.}
\end{eqnarray}
Following the suggestion proposed in Ref.~\cite{strong2008}, we introduce an additional set of position-dependent creation and annihilation operators as 
\begin{eqnarray}
\label{E15}
\hat{a}_{\mathrm{A}'}(\omega) = -\frac{1}{\hbar g_{\mathrm{A}'\mathrm{A}'}(\omega)}\sum_{\lambda=e,m} \int d^3{r} \mathbf{d}_{\mathrm{A}'}^{10} \cdot 
\mathbf{G}_\lambda(\mathbf{r}_{\mathrm{A}'},\mathbf{r},\omega) \cdot\hat{\mathbf{f}}_\lambda(\mathbf{r},\omega),
\end{eqnarray}
 with
\begin{eqnarray}
\label{E16}
g_{\mathrm{A}'\mathrm{A}''}(\omega)= \sqrt{\frac{ \mu_0 }{\hbar \pi} \omega^2\mathbf{d}_{\mathrm{A}'}^{10}\cdot \mathrm{Im}\mathbf{G} (\mathbf{r}_{\mathrm{A}'},\mathbf{r}_{\mathrm{A}''}, \omega)\cdot \mathbf{d}_{\mathrm{A}''}^{01}}\,, \quad \mathrm{A}',\mathrm{A}''\in\{\mathrm{A},\mathrm{B}\}
\end{eqnarray}
defined as the atom-field coupling strength. 
Substitution of these into $\hat{H}_{\mathrm{A}'\mathrm{F}}$ and applying the rotating wave approximation yields
\begin{eqnarray}
\label{E17}
\hat{H}_{\mathrm{A}'\mathrm{F}}= \hbar \int^{\infty}_0 d\omega g_{\mathrm{A}'}(\mathbf{r}_{\mathrm{A}'},\omega)\big[\hat{a}_{\mathrm{A}'}(\mathbf{r}_{\mathrm{A}'},\omega)\hat{\sigma}_{\mathrm{A}'}^\dagger+\hat{a}_{\mathrm{A}'}^\dagger(\mathbf{r}_{\mathrm{A}'},\omega) \hat{\sigma}_{\mathrm{A}'}\big].
\end{eqnarray}
($g_{\mathrm{A}'}\!\equiv\!g_{\mathrm{A}'\mathrm{A}'}$). Making use of commutation relations of bosonic variables together with Eqs. \eref{E15}, \eref{E16} and integral relation \eref{E12}, 
the commutation relations of $\hat{a}(\omega)$ and $\hat{a}^\dagger(\omega)$ can be calculated to be
\begin{eqnarray}
\label{E18}
[\hat{a}_{\mathrm{A}'}(\omega),\hat{a}_{\mathrm{A}''}^\dagger(\omega')]=\frac{g_{\mathrm{A}'\mathrm{A}''}^2(\omega)}{g_{\mathrm{A}'}(\omega)
g_{\mathrm{A}''}(\omega)} \delta(\omega-\omega')
\end{eqnarray}
We define single quantum excitation states 
\begin{eqnarray}
\label{E19}
|\mathbf{r}_{\mathrm{A}'},\omega \rangle = \hat{a}_{\mathrm{A}'}^\dagger(\omega) |\lbrace 0 \rbrace \rangle
\end{eqnarray}
which are eigenstates of $\hat{H}_\mathrm{F}$,
\begin{equation}
\label{E20}
\hat{H}_\mathrm{F} |\mathbf{r}_{\mathrm{A}'},\omega \rangle= \hbar \omega |\mathbf{r}_{\mathrm{A}'},\omega \rangle,
\end{equation}
and have the orthogonality relation as 
\begin{eqnarray}
\label{E21}
\langle\mathbf{r}_{\mathrm{A}'},\omega |\mathbf{r}_{\mathrm{A}''},\omega' \rangle =\frac{g_{\mathrm{A}'\mathrm{A}''}^2(\omega)}{g_{\mathrm{A}'}(\omega)
g_{\mathrm{A}''}(\omega)} \delta(\omega-\omega'),
\end{eqnarray}
being evident from Eq.~(\ref{E18}).
These states are orthogonal with respect to frequency $\omega$ but not position $\mathbf{r}$. This reflects the fact that there is a non-zero probability for the photons emitted by atom positioned at $\mathbf{r}_{\mathrm{A}}$ to be reabsorbed by an atom at a different position $\mathbf{r}_{\mathrm{B}}$.

\section{Van der Waals forces}
\label{3}

In agreement with the pioneering idea of Casimir and Polder~\cite{CP}, two atoms that are weakly coupled to the body-assisted electromagnetic field experience a vdW force deduced from an associated vdW potential. This in turn can be interpreted as the position-dependent energy shift due to a perturbation of the energy of the initial atomic state, arising from the atom-field coupling, so that the force is given by
\begin{eqnarray}
\label{E22}
F_{\mathrm{A}(\mathrm{B})}(\mathbf{r}_{\mathrm{A}},\mathbf{r}_\mathrm{B})=-\nabla_{\mathrm{A}(\mathrm{B})} U(\mathbf{r}_{\mathrm{A}},\mathbf{r}_\mathrm{B}),
\end{eqnarray}
where $U(\mathbf{r}_{\mathrm{A}},\mathbf{r}_\mathrm{B})$ is the vdW potential between two atoms $\mathrm{A}$ and $\mathrm{B}$. However, in the case of excited atoms that are strongly coupled to the electromagnetic field, the interaction energy may not be sufficiently small and hence the perturbative approach
may no longer be valid. This is the scenario we consider here. 

For the following, we restrict our attention to the strong-coupling vdW force between two identical two-level atoms. The paradigmatic model of an atom strongly interacting with an electromagnetic field is the JC model~\cite{jc}. This consists of a single two-level atom (or molecule) interacting with a single near-resonant quantized mode of the electromagnetic field of an ideal cavity, and predicts an oscillatory atom-field excitation exchange.

Here we extend this model to a system comprising two electrically-polarisable atoms interacting with one cavity mode of the body-assisted electromagnetic field, generalizing the simplified picture of standing-wave modes in an ideal cavity. 
 
\subsection{Static approximation}
\label{3.1}

The considered system consists of two identical atoms, which is to say they have the same energy spacing ($\omega_\mathrm{A}^{10}=\omega^{10}_\mathrm{B}\equiv\omega_{10}$). One of the atoms is assumed to be excited, 
and hence, in a strong interaction with a single cavity mode. 
The model is based on the assumption that we may approximate the single cavity mode by a narrow ($\gamma_\nu \ll \omega_\nu$) Lorentzian-type spectrum;
\begin{eqnarray}
\label{E23}
g_{\mathrm{A}'\mathrm{A}''}^2(\omega)= g_{\mathrm{A}'\mathrm{A}''}^2(\omega_\nu)\frac{\gamma_\nu^2/4}{(\omega-\omega_\nu)^2+\gamma_\nu^2/4}\,
\end{eqnarray}
with the parameter $\omega_\nu$ being the central frequency of the peak and $\gamma_\nu$ describing the spectral width.
\\

In the case of two identical atoms, the Hilbert space can be spanned by two coupled states
$ |u_1\rangle\equiv \case{1}{\sqrt{2}}\left(|{1_\mathrm{A}}\rangle\ |{0_\mathrm{B}}\rangle\ +|{0_\mathrm{A}}\rangle\ |{1_\mathrm{B}}\rangle\right) |{0_\nu}\rangle $ 
and 
$|u_2\rangle\equiv |{0_\mathrm{\mathrm{A}}}\rangle\ |{0_\mathrm{\mathrm{B}}}\rangle\  |{1_\nu}\rangle$
with the excitation being shared between atoms ($|{1_{\mathrm{A}'}}\rangle$
) and the field ($|{1_\nu}\rangle$).
It is worth noting that the bare states are chosen according to the fact that the total number of excitations remains constant in the rotating wave approximation.

The state $|{1_\nu}\rangle$ represents an excitation of a single mode $\nu$ and is defined as
\begin{eqnarray}
\label{E24}
|1_\nu\rangle = \sqrt{\frac{\gamma_\nu}{2\pi N}} \sum_{A'=A,B}\int ^{\omega_\nu +\delta \omega_\nu}_{\omega_\nu -\delta \omega_\nu}   d\omega 
\frac{g_{\mathrm{A}'}(\omega_\nu)}{\sqrt{(\omega-\omega_\nu)^2+\gamma_\nu^2/4}} \hspace{0.2cm} |\mathbf{r}_{\mathrm{A}'}, \omega \rangle,
\end{eqnarray}
\\
Here, $\delta\omega$ is regarded as the distance between two neighbouring modes, which in our single-mode assumption must always be much larger than the width $\gamma_\nu$ of that mode.  The quantity $N$ is the normalization factor. 
Note that according to the definition of Lorentzian functions, as
 $\gamma_\nu$ tends to zero, $g_{\mathrm{A}'\mathrm{A}''}^2(\omega_\nu)$ tends to infinity 
 in a way the product of this with a Lorentzian takes the following form; 
\begin{eqnarray}
\label{E25}
\lim_{\gamma_\nu \to 0} g_{\mathrm{A}'\mathrm{A}''}^2(\omega)= \frac{1}{2}\pi\gamma_\nu g_{\mathrm{A}'\mathrm{A}''}^2(\omega_\nu) \delta(\omega-\omega_\nu).
\end{eqnarray}
This fact allows the states $|1_\nu\rangle$ to become normalized to unity. The explicit normalization factor $N$ can be calculated using Eqs. (\ref{E21}) and (\ref{E23}) together with (\ref{E25}), and is found to be
\begin{equation}
\label{E26}
N=g_{\mathrm{A}}^2(\omega_\nu)+g_\mathrm{B}^2(\omega_\nu) +2 g_{\mathrm{AB}}^2(\omega_\nu).
\end{equation}
For the following it is appropriate to construct a dressed basis. Using Eqs. (\ref{E1}) together with Eq. (\ref{E3}), (\ref{E4}) and (\ref{E17}) and the definitions (\ref{E24}) and (\ref{E26}), one can obtain the matrix representation of $\hat{H}$ on the subspace spanned by two states $|u_1\rangle$ and $|u_2\rangle$ as
\begin{equation}
\label{E27}
\hat{H}=\hbar
\left(\begin{array}{cc} 0 & \frac{1}{2}{\Omega_R(\mathbf{r}_\mathrm{A},\mathbf{r}_\mathrm{B})}\\
 \frac{1}{2}{\Omega_R(\mathbf{r}_\mathrm{A},\mathbf{r}_\mathrm{B})}& \Delta \end{array}\right)
\end{equation}
with
\begin{equation}
\label{E28}
\Omega_R(\mathbf{r}_\mathrm{A},\mathbf{r}_\mathrm{B})=\sqrt{\gamma_\nu \pi N}, \qquad \Delta=\omega_\nu -\omega_{10}
\end{equation}
being the vacuum Rabi frequency and detuning, respectively. Diagonalising the Hamiltonian  (\ref{E27}) yields the two eigenvalues
\begin{equation}
\label{E29}
E_\pm =\frac{\hbar}{2}\Delta\pm \frac{\hbar}{2} \Omega(\mathbf{r}_\mathrm{A},\mathbf{r}_\mathrm{B}),
\end{equation}
with $\Omega(\mathbf{r}_\mathrm{A},\mathbf{r}_\mathrm{B})=\sqrt{\Omega^2_R(\mathbf{r}_\mathrm{A},\mathbf{r}_\mathrm{B})+\Delta^2}$.

Finally, the respective dressed states of the combined atom-field system can be written in the form
\begin{eqnarray}
\label{E30}
\vert+\rangle=\cos[\theta_c(\mathbf{r}_{\mathrm{A}},\mathbf{r}_\mathrm{B})] \vert u_1\rangle+ \sin [\theta_c(\mathbf{r}_{\mathrm{A}},\mathbf{r}_\mathrm{B})] \vert u_2\rangle\\
\label{E31}
\vert-\rangle=-\sin[\theta_c(\mathbf{r}_{\mathrm{A}},\mathbf{r}_\mathrm{B})] \vert u_1\rangle+ \cos [\theta_c(\mathbf{r}_{\mathrm{A}},\mathbf{r}_\mathrm{B})] \vert u_2\rangle
\end{eqnarray}
where $\theta_c(\mathbf{r}_{\mathrm{A}},\mathbf{r}_\mathrm{B})$ is the coupling angle defined by
\begin{equation}
\label{E32}
\tan(2\theta_c)=-\frac{\Omega_R(\mathbf{r}_A,\mathbf{r}_B)}{\Delta}\,,
\end{equation}
\begin{equation}
\label{E33}
\cos \theta_c=\frac{\Omega_R(\mathbf{r}_{\mathrm{A}},\mathbf{r}_\mathrm{B})}{\sqrt{\Omega^2_R(\mathbf{r}_{\mathrm{A}},\mathbf{r}_\mathrm{B})+(\Delta+\Omega(\mathbf{r}_{\mathrm{A}},\mathbf{r}_\mathrm{B}))^2}}\,, 
\end{equation}
\begin{equation}
\label{E34}
\sin \theta_c=\frac{\Delta+\Omega(\mathbf{r}_{\mathrm{A}},\mathbf{r}_\mathrm{B})}{\sqrt{\Omega^2_R(\mathbf{r}_{\mathrm{A}},\mathbf{r}_\mathrm{B})+(\Delta+\Omega(\mathbf{r}_{\mathrm{A}},\mathbf{r}_\mathrm{B}))^2}}.
\end{equation}
where we have abbreviated $\theta_c$ $\!=\theta_c(\mathbf{r}_{\mathrm{A}},\mathbf{r}_\mathrm{B})$.

The  position-dependent parts of the eigenenergies can be regarded as dispersion potentials
\begin{eqnarray}
\label{E35}
U_\pm =\pm \frac{\hbar}{2} \Omega(\mathbf{r}_{\mathrm{A}},\mathbf{r}_\mathrm{B}) ,
\end{eqnarray}

The total dispersion force on atom $\mathrm{A}$ consists of a single-atom CP force and the cavity-enhanced vdW force  
from atom $\mathrm{B}$. The latter is defined as;
\begin{eqnarray}
\label{E36}
F^\mathrm{A}_\pm(\mathbf{r}_{\mathrm{A}},\mathbf{r}_\mathrm{B})=-\nabla_{\!\mathrm{A}} U_\pm(\mathbf{r}_{\mathrm{A}},\mathbf{r}_\mathrm{B})
\end{eqnarray}
In order to obtain a more general expression for the force, let us consider the system to be prepared in a superposition state $\vert \theta \rangle$ which has projections on both eigenstates $\vert + \rangle$ and $\vert - \rangle$
\begin{equation}
\label{E37}
\vert \theta \rangle=\cos\theta \vert u_1\rangle+ \sin \theta \vert u_2\rangle=
 \cos(\theta-\theta_c) \vert+\rangle+ \sin(\theta-\theta_c) \vert-\rangle,
\end{equation}
where the last equality is written according to the definitions (\ref{E30}) and (\ref{E31}).
Making use of Eqs. (\ref{E35}) and (\ref{E37}), the potential of the system prepared in the state $\theta$ can be calculated as
\begin{eqnarray}
\label{E38}
&&U_\theta = \vert\langle \theta\vert+\rangle\vert ^2 U_+ + \vert\langle \theta\vert-\rangle\vert ^2 U_- 
= \cos^2(\theta-\theta_c) U_+ + \sin^2(\theta-\theta_c) U_- \nonumber\\
&&\hspace{0.5cm}=\frac{\hbar\Omega}{2} \cos[2(\theta-\theta_c)]
\end{eqnarray}
[$\Omega$ $\!= \Omega(\mathbf{r}_{\mathrm{A}},\mathbf{r}_\mathrm{B}) $], where the position-dependence arises from $\theta_c$ as well as $\Omega$, so the vdW force on atom $\mathrm{A}$ can be evaluated as
\begin{equation}
\label{E39}
F_\theta^\mathrm{A}(\mathbf{r}_{\mathrm{A}},\mathbf{r}_\mathrm{B})=
-\frac{\hbar}{2} \cos [2(\theta-\theta_c)] \nabla_{\!\mathrm{A}} 
\Omega  -\hbar\sin [2(\theta-\theta_c)] \Omega \nabla_{\!\mathrm{A}} \theta_c.
\end{equation}
 This equation can be simplified by performing the differentiations in the right hand side.
 Using Eq. (\ref{E32}) and the identity
 $(1+\tan^2\beta)^{1/2}$ $\!=\cos \beta$, we find
\begin{eqnarray}
\label{E40}
\nabla_{\!\mathrm{A}} \theta_c=-\frac{1}{2\Delta}\cos^2(2\theta_c)\nabla_{\!\mathrm{A}} \Omega_R 
\end{eqnarray}
[$\Omega_R$ $\!= \Omega_R(\mathbf{r}_{\mathrm{A}},\mathbf{r}_\mathrm{B}) $], using which, together with Eq. (\ref{E34}), we obtain 
$\sin(2\theta_c)=\Omega_R/\Omega$.
Finally, using the definition of $\Omega(\mathbf{r}_{A},\mathbf{r}_B)$, we find
\begin{equation}
\label{E41}
\nabla_{\!\mathrm{A}} \Omega = \sin(2\theta_c)
\nabla_{\!\mathrm{A}} \Omega_R.
\end{equation}
Substituting this result into Eq. (\ref{E39}), the vdW force on atom $\mathrm{A}$, prepared in superposition state $\theta$,
can be written as
\begin{eqnarray}
\label{E42}
&& \hspace{-.8in}F_\theta^\mathrm{A}(\mathbf{r}_{\mathrm{A}},\mathbf{r}_\mathrm{B})=-\frac{\hbar}{2}  
\left\{ \cos [2(\theta-\theta_c)] +\cot(2\theta_c) \sin [2(\theta-\theta_c)] \right\}\nabla_{\!\mathrm{A}} \Omega_R\nonumber\\
&&
=-\frac{\hbar}{2} \frac{\sin\theta\cos\theta}{\sin\theta_c\cos\theta_c}\nabla_{\!\mathrm{A}} \Omega_R.
\end{eqnarray}

Let us study the state dependence of the force by considering special cases. It is clear from the above formula that if $\theta=\theta_c$ or $\theta=\theta_c+\pi/2$ (which is equivalent to $|\theta \rangle=|+\rangle$ or $|\theta \rangle=|-\rangle$, respectively), the previous results of Eq. (\ref{E36}) can be recovered and if $\theta=0$ ($|\theta\rangle=|u_1\rangle$) or $\theta=\pi/2$ ($|\theta\rangle=|u_2\rangle$), the vdW force will vanish. We will study the weak coupling limits of the force in next section.

\subsection{Weak coupling}
\label{3.2}

Let us return to the primary idea of Casimir and Polder and consider a continuous and unbounded spectrum for the field. According to the perturbation theory for two identical atoms, the lowest-order energy shift depending on the positions of both atoms is second order and is defined as
\begin{eqnarray}
\label{E43}
\bigtriangleup E = -\sum_{I}\sum_{\mathrm{A}'=\mathrm{A},\mathrm{B}}\frac{\langle \psi| \hat{H}_{\mathrm{A}'\mathrm{F}}|I\rangle \langle I|\hat{H}_{\mathrm{A}'\mathrm{F}}| \psi \rangle}
{E_{I}-E_{\psi}}
\end{eqnarray}
in which $| \psi \rangle$ and $|I\rangle $ are the initial and intermediate states, respectively, and $E_{\psi}$ and $E_{I}$ are their respective unperturbed eigenenergies. We will consider $|\psi \rangle$ $\!= \case{1}{\sqrt{2}}\left(|{1_\mathrm{A}}\rangle |{0_\mathrm{B}}\rangle\right.$ $\left.\!+|{0_\mathrm{A}}\rangle |{1_\mathrm{B}}\rangle \right) |\lbrace0\rbrace\rangle $ and $|I\rangle=|0_\mathrm{A}\rangle |0_\mathrm{B}\rangle|\mathbf{1}_\lambda (\mathbf{r},\omega)\rangle$, with $|\mathbf{1}_\lambda (\mathbf{r},\omega)\rangle$ $\!=\hat{f}_\lambda^\dagger (\mathbf{r},\omega)|\lbrace0\rbrace\rangle $.

Note that the particular choice of $|I\rangle$ ensures that we only deal with a real photon exchange process, so the potential will be resonant and the sum over intermediate state in Eq. (\ref{E43}) is in fact the sum over polarizations $\lambda$ and integrals over $\mathbf{r}$ and $\omega$. 

Using Eqs. (\ref{E7}), (\ref{E9}) and (\ref{E13}) together with Cauchy's theorem, the resonant potential for two identical atoms (one of them being excited) that are weakly coupled to body-assisted field, reads
\begin{eqnarray}
\label{E44}
&&U=-\frac{1}{2}\mu_0 \omega_{10}^2 \mathbf{d}_\mathrm{A}^{10}\cdot \mathrm{Re}\mathbf{G} (\mathbf{r}_\mathrm{A},\mathbf{r}_\mathrm{A}, \omega_{10})\cdot\mathbf{d}_\mathrm{A}^{01}\nonumber\\
&&\hspace{0.8cm}-\frac{1}{2}\mu_0 \omega_{10}^2 \mathbf{d}_\mathrm{B}^{10}\cdot \mathrm{Re}\mathbf{G} (\mathbf{r}_\mathrm{B},\mathbf{r}_\mathrm{B}, \omega_{10})\cdot\mathbf{d}_\mathrm{B}^{01}\nonumber\\
&&\hspace{0.8cm}-\mu_0 \omega_{10}^2 \mathbf{d}_\mathrm{A}^{10}\cdot \mathrm{Re}\mathbf{G} (\mathbf{r}_\mathrm{A},\mathbf{r}_\mathrm{B}, \omega_{10})\cdot\mathbf{d}_\mathrm{B}^{01} .
\end{eqnarray}

The first and second terms represent the single-atom potentials for atom $\mathrm{A}$ and $\mathrm{B}$, respectively, while the third is the interaction potential of the two atoms. To make contact with the perturbative results obtained by McLone and Power \cite{identical1964} for two identical atoms in free space, let us consider the Green function in free space as~\cite{novotny2012Greentensor}
\\
\begin{equation}
\label{E45}
G_{\alpha\beta}(k,\mathbf{r})=\frac{e^{ikr}}{4 \pi r} \left[\left(1+\frac{ikr-1}{k^2r^2}\right)\delta_{\alpha\beta}+\left(-1+\frac{3-3ikr}{k^2r^2}\right)\frac{r_{\alpha}r_\beta}{r^2}\right],
\end{equation}
where $k={\omega}/{c}$, $\mathbf{r}=\mathbf{r}_2-\mathbf{r}_1$, $r=|\mathbf{r}|$, $r_{\alpha}=\mathbf{e}_{\alpha}\cdot \mathbf{r}$.
Substituting the real part of this tensor in the third part of Eq. (\ref{E46}), leads to
\begin{equation}
\label{E46}
\fl 
U(r)=-d_{A\alpha} d_{B\beta} \left[(\delta_{\alpha\beta}-\mathbf{e}_{\alpha}\mathbf{e}_\beta)\frac{k^2 \cos(kr)}{r}-(\delta_{\alpha\beta}-3\mathbf{e}_{\alpha}\mathbf{e}_\beta)\left(\frac{k\sin(kr)}{r^2}+\frac{\cos(kr)}{r^3}\right)\right]
\end{equation}
in agreement with previous results of Ref.~\cite{identical1964}.

To make contact with our method, we consider the limiting case of weak coupling in Eq. (\ref{E35}). In this limit $\Delta\gg \Omega_R(\mathbf{r}_A,\mathbf{r}_B)$, which means that $\theta_c=\pi/2$, $\vert+\rangle=\vert u_2 \rangle $ and 
$\vert-\rangle=-\vert u_1 \rangle$. The potential associated with states $|u_1\rangle $ and $|u_2\rangle $ respectively can then be written as
\begin{eqnarray}
\label{E47}
&&U_- =- \frac{\hbar \gamma_\nu \pi N}{4\Delta} 
\end{eqnarray} 
\begin{eqnarray}
\label{E48}
&&U_+ = \frac{\hbar\gamma_\nu \pi N}{4\Delta} 
\end{eqnarray} 
Employing the Kramers-Kronig relation
\begin{eqnarray}
\label{E49}
&&\omega^2 \mathrm{Re} \mathbf{G}(\mathbf{r}_{\mathrm{A}},\mathbf{r}_\mathrm{B},\omega)=\frac{1}{\pi} \mathrm{P} \int_{-\infty}^{\infty} \frac{d\omega}{\omega'-\omega} \omega'^2 \mathrm{Im}\mathbf{G}(\mathbf{r}_{\mathrm{A}},\mathbf{r}_\mathrm{B},\omega'),
\end{eqnarray} 
for a sufficiently narrow mode (recalling Eqs. (\ref{E23}) and (\ref{E25})) we find:
\begin{eqnarray}
\label{E50}
\fl  &&\mu_0 \omega^2 \mathbf{d}_{\mathrm{A}'}^{10}\cdot \mathrm{Re} \mathbf{G}^{\mathrm{L}} ( \mathbf{r}_{\mathrm{A}'} , \mathbf{r}_{\mathrm{A}''} , \omega)\cdot \mathbf{d}_{\mathrm{A}''}^{01} \nonumber\\
&&\hspace{0.2cm}= \frac{\mu_0}{\pi} \omega_\nu^2 \mathbf{d}_{\mathrm{A}'}^{10}\cdot \mathrm{Im} \mathbf{G}^{\mathrm{L}} ( \mathbf{r}_{\mathrm{A}'} , \mathbf{r}_{\mathrm{A}''} , \omega_\nu)\cdot \mathbf{d}_{\mathrm{A}''}^{10} \mathrm{P} \int_{-\infty}^\infty \frac{d\omega}{\omega'-\omega}
\frac{\gamma_\nu^2/4}{(\omega'-\omega_\nu)^2+\gamma_\nu^2/4} \nonumber\\
&&\hspace{0.2cm}=\frac{\mu_0 \gamma_\nu}{2(\omega_\nu-\omega)}\omega_\nu^2 \mathbf{d}_{\mathrm{A}'}^{10}\cdot \mathrm{Im} \mathbf{G}^{\mathrm{L}} ( \mathbf{r}_{\mathrm{A}'} , \mathbf{r}_{\mathrm{A}''} , \omega_\nu)\cdot \mathbf{d}_{\mathrm{A}''}^{10},
\end{eqnarray}
Note that the sign $\mathrm{L}$ has been added to indicate the assumption of dealing with single-resonance, Lorentzian cavity-field.

Using this result, Eq. (\ref{E47}) and (\ref{E48}), respectively, can be given as
\begin{eqnarray}
\label{E51}
&&U_-=-\frac{1}{2}\mu_0 \omega_{10}^2 \mathbf{d}_\mathrm{A}^{10}\cdot \mathrm{Re}\mathbf{G}^{\mathrm{L}} (\mathbf{r}_\mathrm{A},\mathbf{r}_\mathrm{A}, \omega_{10})\cdot\mathbf{d}_\mathrm{A}^{01}\nonumber\\
&&\hspace{1cm}-\frac{1}{2}\mu_0 \omega_{10}^2 \mathbf{d}_\mathrm{B}^{10}\cdot \mathrm{Re}\mathbf{G}^{\mathrm{L}} (\mathbf{r}_\mathrm{B},\mathbf{r}_\mathrm{B}, \omega_{10})\cdot\mathbf{d}_\mathrm{B}^{01}\nonumber\\
&&\hspace{1cm}-\mu_0 \omega_{10}^2 \mathbf{d}_\mathrm{A}^{10}\cdot \mathrm{Re}\mathbf{G}^{\mathrm{L}} (\mathbf{r}_\mathrm{A},\mathbf{r}_\mathrm{B}, \omega_{10})\cdot\mathbf{d}_\mathrm{B}^{01} 
\end{eqnarray}
and 
\begin{eqnarray}
\label{E52}
&&U_+=- U_- .
\end{eqnarray}
\Eref{E51} is in complete agreement with the time-independent perturbative results for the resonant part of the dispersion potential of two identical two-level atoms (one of them being prepared in the upper state) that are weakly coupled to a body-assisted electromagnetic field, while Eq. (\ref{E52}) is the weak coupling potential for the state with excited field and ground state atoms, and has not been predicted by perturbation theory so far.

In complete analogy to the derivation of Eqs. (\ref{E51}) and (\ref{E52}), one can show that in weak coupling limit (large detuning)the potential and the corresponding force on one of the atoms, for a system prepared in state $\vert\theta\rangle$, can be written as
\begin{eqnarray}
\label{E53}
U_\theta(\mathbf{r}_{\mathrm{A}},\mathbf{r}_\mathrm{B})=-\frac{\hbar}{4\Delta} \cos(2\theta)\Omega^2_R(\mathbf{r}_{\mathrm{A}},\mathbf{r}_\mathrm{B})\nonumber\\
\hspace{2cm}=-\frac{1}{2}\cos(2\theta)\mu_0 \omega_{10}^2 \mathbf{d}_\mathrm{A}^{10}\cdot \mathrm{Re}\mathbf{G}^{\mathrm{L}} (\mathbf{r}_\mathrm{A},\mathbf{r}_\mathrm{A}, \omega_{10})\cdot\mathbf{d}_\mathrm{A}^{01}\nonumber\\
\hspace{2.4cm}-\frac{1}{2}\cos(2\theta)\mu_0 \omega_{10}^2 \mathbf{d}_\mathrm{B}^{10}\cdot \mathrm{Re}\mathbf{G}^{\mathrm{L}} (\mathbf{r}_\mathrm{B},\mathbf{r}_\mathrm{B}, \omega_{10})\cdot\mathbf{d}_\mathrm{B}^{01}\nonumber\\
\hspace{2.4cm}-\cos(2\theta)\mu_0 \omega_{10}^2 \mathbf{d}_\mathrm{A}^{10}\cdot \mathrm{Re}\mathbf{G}^{\mathrm{L}} (\mathbf{r}_\mathrm{A},\mathbf{r}_\mathrm{B}, \omega_{10})\cdot\mathbf{d}_\mathrm{B}^{01}
\end{eqnarray}
and
\begin{eqnarray}
\label{E54}
F_\theta(\mathbf{r}_{\mathrm{A}})=\frac{\hbar}{2\Delta} \cos(2\theta)\Omega_R(\mathbf{r}_{\mathrm{A}},\mathbf{r}_\mathrm{B}) \nabla_{\!\mathrm{A}}\Omega_R(\mathbf{r}_{\mathrm{A}},\mathbf{r}_\mathrm{B}).
\end{eqnarray}

\section{Planar cavity}
\label{4}

\begin{figure}[h]
\includegraphics[scale=0.4,center]{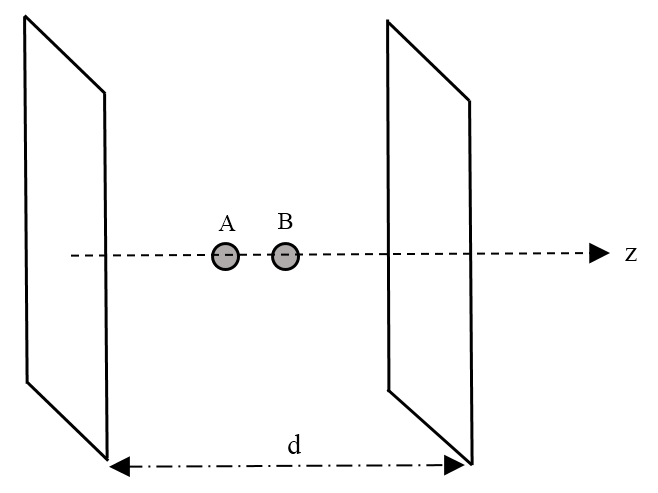}
\caption{Two atoms in a planar cavity.}
\label{F1}
\end{figure}

To apply our model, we assume two atoms, $\mathrm{A}$ and $\mathrm{B}$, to be placed in a planar cavity of width $d$. The cavity plates are assumed to be identical and -almost- perfectly reflecting and the atoms are placed along an axis perpendicular to the cavity plates ($z$ direction), see Fig. \ref{F1}.

We use the Green's tensor introduced in Ref.~\cite{buhbookI};
\begin{eqnarray}
\label{E55}
&\fl \mathbf{G}(\mathbf{r},\mathbf{r}',\omega)= \frac{i \mu(\omega)}{8\pi} \int \frac{d^2 k^\Vert}{k^  \bot} e^{i k^\Vert \cdot (\mathbf{r}-\mathbf{r}')} \nonumber\\
&\sum_{\sigma=s,p}\Big\lbrace \frac{r_\sigma^2 e^{2ik^\bot d}}{D_\sigma}\Big[\mathbf{e}_{\sigma +}\mathbf{e}_{\sigma +} e^{i k^\bot(z-z')}+\mathbf{e}_{\sigma -}\mathbf{e}_{\sigma -} e^{-i k^\bot(z-z')}\Big]\nonumber\\
&+\frac{1}{D_\sigma}\Big[\mathbf{e}_{\sigma +}\mathbf{e}_{\sigma -} r_\sigma e^{i k^\bot(z+z')}+\mathbf{e}_{\sigma -}\mathbf{e}_{\sigma +} r_\sigma e^{-i k^\bot(2d-z-z')}\Big]\Big\rbrace ,
\end{eqnarray}
with
\begin{eqnarray}
\label{E56}
 D_\sigma = 1-r_\sigma^2 e^{2id k^\bot}
\end{eqnarray}
\begin{eqnarray}
\label{E57}
k^\bot = \sqrt{k^2- {k^\Vert}^2} .
\end{eqnarray}
$\mathbf{e}_{s\pm}$ and $\mathbf{e}_{p\pm}$ are the unit vectors for $s$/$p$-polarised waves.
The reflection coefficients are $r_p=-r_s=1-\delta$, in which
$\delta$, regarded as a small deviation of the plates from being perfectly reflecting, can be related to the transmission coefficient of the plates through $\delta=t^2/2$.

On cavity resonances ($\omega_\nu=\nu \pi c/d$), $k^\Vert\simeq0$ carries the main contribution to the integral.
Taking one parallel component of Green's tensor and carrying out the angular integral, we get to a simplified version of respective Green's tensor, by which we find:
\begin{eqnarray}
\label{E58}
&\fl \omega_\nu^2 \mathrm{Im}G_{xx}(z_\mathrm{A},z_\mathrm{B},\omega_\nu)= -\frac{c\omega_\nu^3}{16 \pi c^2 \delta} \Bigg\lbrace\cos{\Bigg[\frac{(2d-z_\mathrm{A}-z_\mathrm{B})\omega_\nu}{c}\Bigg]}-\cos{\Bigg[\frac{(2d+z_\mathrm{A}-z_\mathrm{B})\omega_\nu}{c}\Bigg]}\nonumber\\
& \hspace{1.5cm} -\cos{\Bigg[\frac{(2d-z_\mathrm{A}+z_\mathrm{B})\omega_\nu}{c}\Bigg]}-\cos{\Bigg[\frac{(z_\mathrm{A}+z_\mathrm{B})\omega_\nu}{c}\Bigg]}\Bigg\rbrace .
\end{eqnarray}
It can be shown that Eq. (\ref{E58}), shows Lorentzian behaviour Eq. (\ref{E23}), in the vicinity of each cavity resonance with $\gamma_\nu=2c\delta/d$.

According to Eqs. (\ref{E26}) and (\ref{E28}), in the case of exact resonance ($\Delta=0$), the cavity-induced Rabi frequencies can be written as
\begin{eqnarray}
\label{E59}
 &\Omega^2= \Omega_R^2=\gamma_\nu \pi \Big[g^2_\mathrm{A}(\omega_\nu) +g^2_\mathrm{B}(\omega_\nu) +2 g^2_\mathrm{AB}(\omega_\nu)\Big]\nonumber\\
&\hspace{1.5cm} \equiv \Omega_A^2 (\mathbf{r}_\mathrm{A}) +\Omega_B^2 (\mathbf{r}_\mathrm{B}) + \Omega_{\mathrm{AB}}^2(\mathbf{r}_\mathrm{A},\mathbf{r}_\mathrm{B}),
\end{eqnarray}
where using the definition for $\gamma_\nu$ together with Eqs. (\ref{E16}) and (\ref{E58}), $\Omega_{\mathrm{AB}}^2(\mathbf{r}_\mathrm{A},\mathbf{r}_\mathrm{B}) \equiv \Omega_{\mathrm{AB}}^2$ is calculated to be
\begin{eqnarray}
\label{E60}
& \Omega_{\mathrm{AB}}^2= \frac{3c \Gamma_0}{4d}\Bigg\lbrace\cos{\Bigg[\frac{(2d-z_\mathrm{A}-z_\mathrm{B})\omega_\nu}{c}\Bigg]}-\cos{\Bigg[\frac{(2d+z_\mathrm{A}-z_\mathrm{B})\omega_\nu}{c}\Bigg]}\nonumber\\
&\hspace{1cm}-\cos{\Bigg[\frac{(2d-z_\mathrm{A}+z_\mathrm{B})\omega_\nu}{c}\Bigg]}-\cos{\Bigg[\frac{(z_\mathrm{A}+z_\mathrm{B})\omega_\nu}{c}\Bigg]}\Bigg\rbrace ,
\end{eqnarray}
with
\begin{eqnarray}
\label{E61}
\Gamma_0=\frac{\omega_{10}^3 |d_{10}|^2}{3 \pi \varepsilon_0 \hbar c^3}
\end{eqnarray}
being the free space decay rate.
Note that $\Omega_\mathrm{A}^2$ and $\Omega_\mathrm{B}^2$ can be calculated from Eq. (\ref{E60}) as the limiting case of $z_\mathrm{A}=z_\mathrm{B}$, divided by 2.

\begin{figure}[h]
\includegraphics[scale=0.5,center]{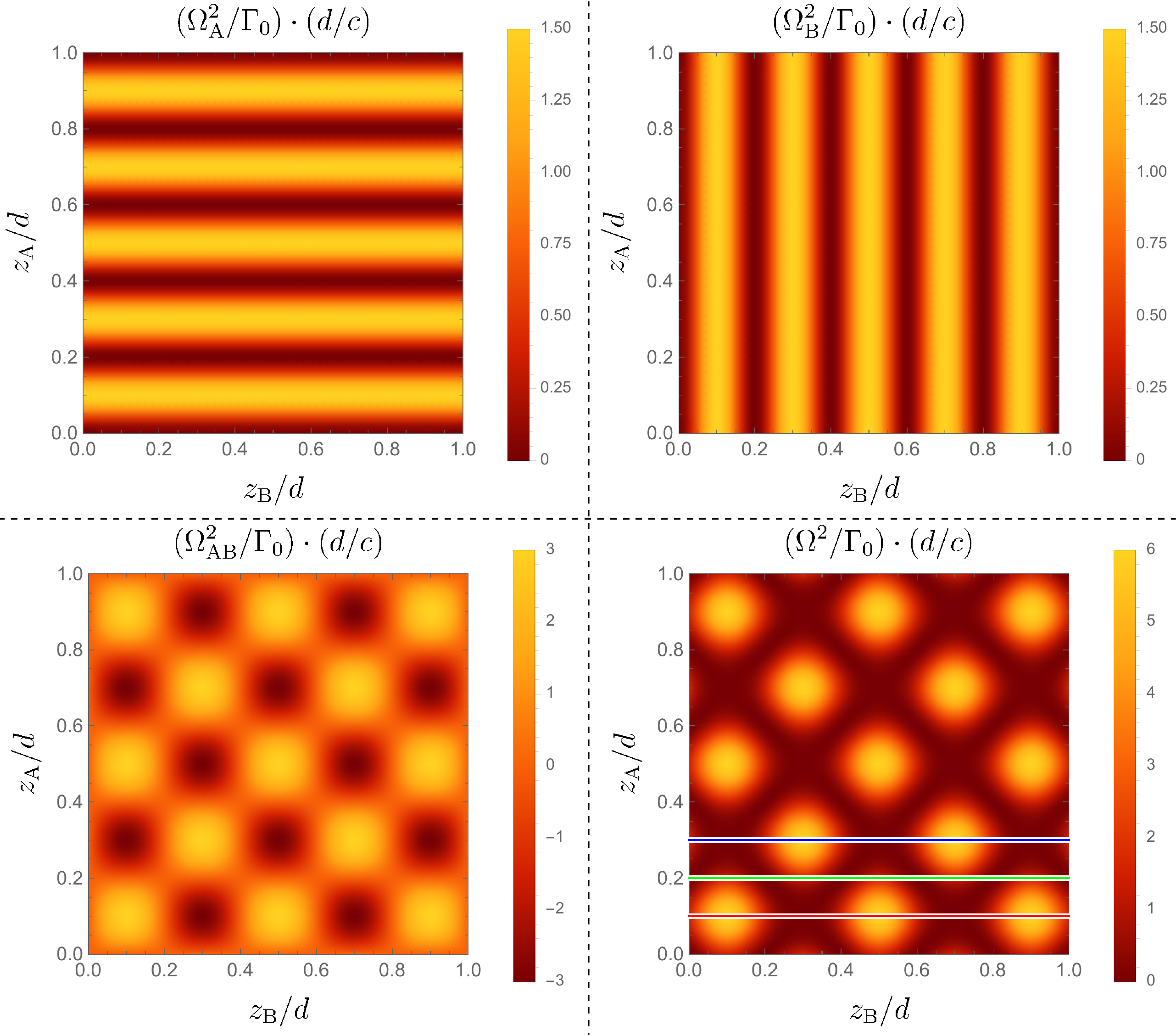}
\caption{Position dependence of dimensionless squared Rabi frequency contributions as well as the total squared Rabi frequency.}
\label{F2}
\end{figure}

\begin{figure}[h]
\includegraphics[scale=0.6,center]{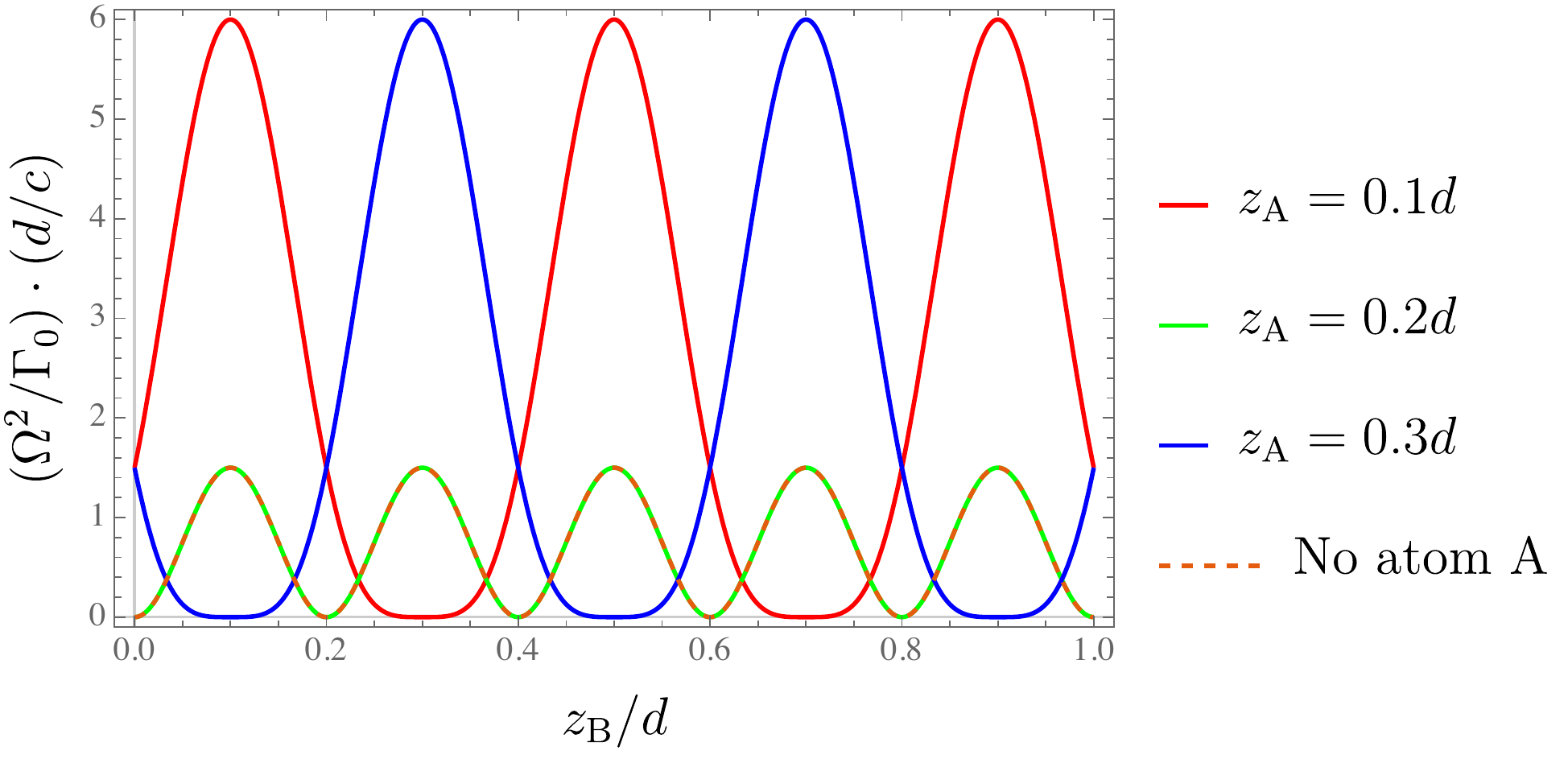}
\caption{The dependence of the squared Rabi frequency on the position of atom $\mathrm{B}$ for given positions of atom $\mathrm{A}$.}
\label{F3}
\end{figure}

The position-dependence of the dimensionless squared Rabi frequencies is shown in Fig. \ref{F2} and Rabi oscillations are observed for the $\Omega_{\mathrm{A/B}}^2$, $\Omega_{\mathrm{AB}}^2$ and the total $\Omega^2$.
It can be seen that, as the two atoms are identical, their contribution to total squared Rabi frequency has the same position-dependence and also the same intensity, while the peaks of $\Omega^2$ are placed in the same position with its interaction contribution but with higher intensity. The dependence of the squared Rabi frequency on the position of atom $\mathrm{B}$ for different positions of atom $\mathrm{A}$ is given in Fig. \ref{F3}. It shows that, if  atom $\mathrm{A}$ is positioned at one of the cavity nodes, the curve coincides with the case where there is only atom $\mathrm{B}$ in the cavity and the presence of the other atom may have no effect on total Rabi oscillations, while, in the case of putting it on one of the anti-nodes, we see some peaks that are roughly four times higher than the case of having only one atom. Different positions for atom $\mathrm{A}$ in Fig. \ref{F3} correspond to the coloured lines in the total squared Rabi frequency shown in Fig. \ref{F2}.  

\section{Summary}
\label{5}
On the basis of macroscopic cavity-QED and using an extended Jaynes-Cummings model, we have obtained general expressions for the van der Waals potential between two atoms and also for the van der Waals force acting on them. The system under consideration consists of two identical two-level atoms that are strongly coupled to one cavity mode. The spectral structure of the mode has been chosen to be Lorentzian and as our single mode assumption required, the mode was assumed to dominate the field in its vicinity. A correlated state basis has been considered and the position-dependent part of the two respective eigenvalues has been evaluated as dispersion potentials of the atoms which only differ by a sign for the two eigenstates. An oscillatory energy exchange between the atomic system and the field (at rate $\Omega$) was predicted.

We have shown that our general expression can be reduced to perturbative results in the appropriate limit of weak coupling. For a system prepared in one the eigenstates, these limits of the potential have been shown to give the resonant perturbative potential of two identical atoms (with one of them being excited) interacting with a field in the vacuum state. The result related to the other eigenstate (the field being excited, not the atoms), was completely new and has not been predicted by perturbation theory so far.

To apply our model, we studied the Rabi oscillations of an atom-field system  placed in a highly reflective planar cavity. We have shown that the oscillations have the same position-dependence and amplitude for each atom, as it should be since we have identical atoms. Showing the squared Rabi frequency dependence on the position of one atom, it has been demonstrated that the second atom would be completely invisible to the first one, if it is positioned at each cavity nodes. Conversely, it may have the equal contribution as the first one, if it is positioned at one of the anti-nodes.

\section*{Acknowledgements}
The authors acknowledge fruitful discussions with Sebastian Fuchs. S.E. acknowledges support from the Ministry of Science, Research and Technology of Iran (MSRT) and the German Academic Exchange Service (DAAD). R.B. acknowledges the Alexander von Humboldt Foundation. S.Y.B. acknowledges support by the Deutsche Forschungsgemeinschaft (grant BU 1803/3-1476), and the Freiburg Institute for Advanced Studies.

\end{document}